\documentclass[preprint]{aastex63}

\DeclareUnicodeCharacter{2212}{-}

\shorttitle{Two Successive Metric Type II Radio Bursts}
\shortauthors{Ramesh et al.}
\graphicspath{{./}{figures/}}
\usepackage{bm}

\begin{document}

\title{Solar coronal density turbulence and magnetic field strength at the source regions of two successive metric type II radio bursts}

\correspondingauthor{R. Ramesh}
\email{ramesh@iiap.res.in}

\author{R. Ramesh}
\affiliation{Indian Institute of Astrophysics, Koramangala 2nd Block, Bangalore 560034, Karnataka, India}

\author{C. Kathiravan}
\affiliation{Indian Institute of Astrophysics, Koramangala 2nd Block, Bangalore 560034, Karnataka, India}

\author[0000-0001-5742-9033]{Anshu Kumari}
\affiliation{Department of Physics, University of Helsinki, P.O.Box 64, 00014 Helsinki, Finland}

\begin{abstract}

We report spectral and polarimeter observations of two weak, low frequency 
(${\approx}$85-60\,MHz)
solar coronal type II radio bursts that occurred on 2020 May 29 within a time interval ${\approx}$2\,min.
The bursts had fine structures, and were due to harmonic plasma emission. Our analysis indicates that the magnetohydrodynamic (MHD) shocks responsible for the 1st and 2nd type II bursts were generated by the leading edge (LE) of an extreme-ultraviolet (EUV) flux rope/coronal mass ejection (CME) and interaction of its flank with a neighbouring coronal structure, respectively. 
The CME deflected from the radial direction by 
${\approx}25^{\arcdeg}$ during propagation in the near-Sun corona.
The estimated power spectral density (PSD) and magnetic field strength ($B$) near the location of the 1st burst at heliocentric distance $r{\approx}1.35R_{\odot}$ are 
$\rm {\approx}2{\times}10^{-3}\,W^{2}m$ and ${\approx}$1.8\,G, respectively. The corresponding values for the 2nd burst at the same $r$ are 
$\rm {\approx}10^{-3}\,W^{2}m$ and ${\approx}$0.9\,G. The significant spatial scales of the coronal turbulence at the location of the two type II bursts are  
${\approx}$62\,-\,1\,Mm.
Our conclusions from the present work are that the turbulence and magnetic field strength in the coronal region near the CME LE are higher compared to the corresponding values close to its flank. The derived estimates of the two parameters correspond to the same $r$ for both the CME LE and its flank, with a delay of ${\approx}$2\,min for the latter.

\end{abstract}

\keywords{Sun: activity; Sun: corona; Sun: coronal mass ejections: CMEs; Sun: radio radiation}

\section{Introduction} \label{sec:intro}

Solar type II radio bursts appear in the spectrograph records as slowly drifting emission lanes from high to low frequencies. They are due to plasma oscillations caused by the electrons accelerated at the MHD shocks propagating outward in the solar atmosphere. 
These shocks are caused by coronal mass ejections (CMEs) and/or flares. The frequency drift rate ($\sim$0.5\,MHz\,s$^{-1}$) of the bursts result from the decrease of electron density ($N_{e}$) and hence the plasma frequency ($f_{p}$), with increasing $r$.
The detailed characteristics of type II bursts could be found in \cite{Nelson1985,Mann1995,Aurass1997,Gopalswamy2006,Nindos2011}.
Sometimes two type II bursts occur in quick succession within a time interval of 
${\sim}10$\,min. They were first reported by \cite{Robinson1982}. The occurrence of such events are attributed to either two successive flares or two successive CMEs or a flare and CME, or leading edge (LE) and flank of a CME \citep{Mancuso2004,Shanmugaraju2005,Subramanian2006,Cho2008,Cho2011,Hariharan2015,
Lv2017,Koval2021}. The CME driven type II bursts could occur at locations along the front of the shock wherever appropriate conditions for electron acceleration are satisfied \citep{Knock2005,Kouloumvakos2021,Jebaraj2021,Ramesh2022a}. Statistical study using two-dimensional imaging observations of coronal type II bursts observed near the solar limb by \cite{Ramesh2012a} indicate that they are located within the angular range 
${\lesssim}46^{\arcdeg}$ from the central position angle of the LE of the associated CMEs. 

Occasionally type II bursts show fine structure in both time and frequency domains. The bandwidth of emission is related to the size scales of the density inhomogeneities or turbulence in the corona \citep[see e.g.][]{Mugundhan2017}. The observed angular broadening of the `radio' Sun at low frequencies is considered to be due to scattering of radio waves by similar inhomogenities \citep{Sastry1994,Ramesh2006a,Thejappa2008,Zhang2022}. The spatial scales of such inhomogeneities has been recently reported by \cite{Carley2021} using observations of the fine structures in type II bursts. The distribution follows a power law with spectral index in the range -1.7 to -2.0 at $r{\approx}2R_{\odot}$ which is close to the value of $−5/3$ expected of fully developed Kolmogorov-like turbulence. 
Note that the power spectrum analysis mentioned above is carried out by first converting the frequency range of observation to heliocentric distance range using a coronal density model. Then, autocorrelation of the radio flux (which will be a function of heliocentric distance after the aforementioned conversion) and its Fourier transformation are carried out \citep[see e.g.][]{Chen2018}. 
Moving further, it is known that plasma emission in a magnetic field gets split as ordinary ($O$) and extraordinary ($X$) modes. Since the propagation characteristics of these two modes are different, there will be a resultant circular polarization \citep{Melrose1972}. In the case of harmonic plasma emission, the associated $B$ can be estimated in a relatively simple manner (see e.g. \cite{Melrose1980corrected,Zlotnik1981}). Several such estimates of $B$ using observations of weak circularly polarized emission from harmonic type II bursts are there in the literature \citep{Hariharan2014,Anshu2017a,Anshu2019,Ramesh2022b,Ramesh2022c}. 
The above mentioned work by various authors indicate that PSD and/or $B$ are useful parameters to compare successive type II bursts. But our current knowledge is very limited.
For e.g. there are only a few published reports of $B$ at different locations along a coronal shock close to the Sun. Using ultraviolet spectra and white-light observations of a partial `halo' CME in the plane of the sky, \cite{Bemporad2014} showed that $B$ near the LE (flank) of the CME at $r{\approx}$2.6$R_{\odot}$ (2.3$R_{\odot}$) is ${\approx}$0.21\,G (0.24\,G). \cite{Koval2021} reported spectral observations of two `fractured' type II bursts due to the interaction of the nose of a rising CME/shock with a pseudo streamer, and its flank with a flux tube. The estimated $B$ values from the two bursts at 
$r{\approx}$2.6$R_{\odot}$ were ${\approx}$0.8 \& 1\,G, respectively. 
Hence the present work.

\section{Observations} \label{sec:obs}

The radio spectral data were obtained with the
GAuribidanur Pulsar System \citep[GAPS,][]{Kshitij2022} in the Gauribidanur Observatory \citep{Ramesh2011a,Ramesh2014} located about 100\,km north of Bangalore\footnote{https://www.iiap.res.in/?q=centers/radio}. The front-end of GAPS has an one-dimensional array of sixteen log-periodic dipole antennas 
\citep[LPDA,][] {Ramesh1998} set up along a North-South baseline. The frequency range of operation is 
85\,-\,45 MHz. The half-power width of the array response pattern (`beam') for observations near the zenith is 
${\approx}110^{\arcdeg}{\times}3^{\arcdeg}$ (right ascension, R.A.\,{$\times$}\,declination, decl.). The width in the direction of R.A. is frequency independent. Along declination, it is at the highest frequency of operation, 
i.e.\,85\,MHz.
The observations were carried out with a Field Programmable Gate Array (FPGA) based digital back-end receiver 
system \citep{Mugundhan2018b} over the aforementioned frequency 
range with a sampling rate of ${\approx}$90\,MHz. Data acquisition were simultaneous at all the frequencies. The spectral bandwidth and integration time are ${\approx}$44\,kHz and $\approx$4\,msec, respectively \citep[see][]{Kshitij2022}.
For polarization data, we used observations with the Gauribidanur RAdio Spectro-Polarimeter \citep[GRASP,][]{Kishore2015}. It has two LPDAs in orthogonal orientation to each other \citep{Sasi2013a} for observations of Stokes I \& V emission. 
The response pattern of each LPDA is wide with half-power width  
${\approx}80^{\arcdeg}$ in both right ascension and declination, independent of frequency. 
The antenna and the receiver systems are routinely calibrated by carrying
out observations in the direction of the Galactic center as described in \cite{Kishore2015}.  The minimum degree of circular polarization ($dcp{=}|V|/I$) detectable with GRASP is ${\lesssim}$0.01.
Linear polarization from the solar atmosphere is absent at low radio frequencies \citep{Grognard1973,Morosan2022}.
For information on CMEs, we made use of the catalog generated from observations in white-light with the Large Angle and Spectrometric Coronagraph C2 
\citep[LASCO C2,][]{Brueckner1995} onboard the SOlar and Heliospheric Observatory 
(SOHO)\footnote{https://cdaw.gsfc.nasa.gov/CME{\_}list/}.
For information on the associated solar surface activity, 
we 
used data obtained in Extreme Ultra-Violet (EUV) at 193{\AA} with the 
Atmospheric Imaging Assembly \citep[AIA,][]{Lemen2012} on board the Solar Dynamics Observatory (SDO).

\begin{figure}[t!]
\centerline{\includegraphics[width=14cm]{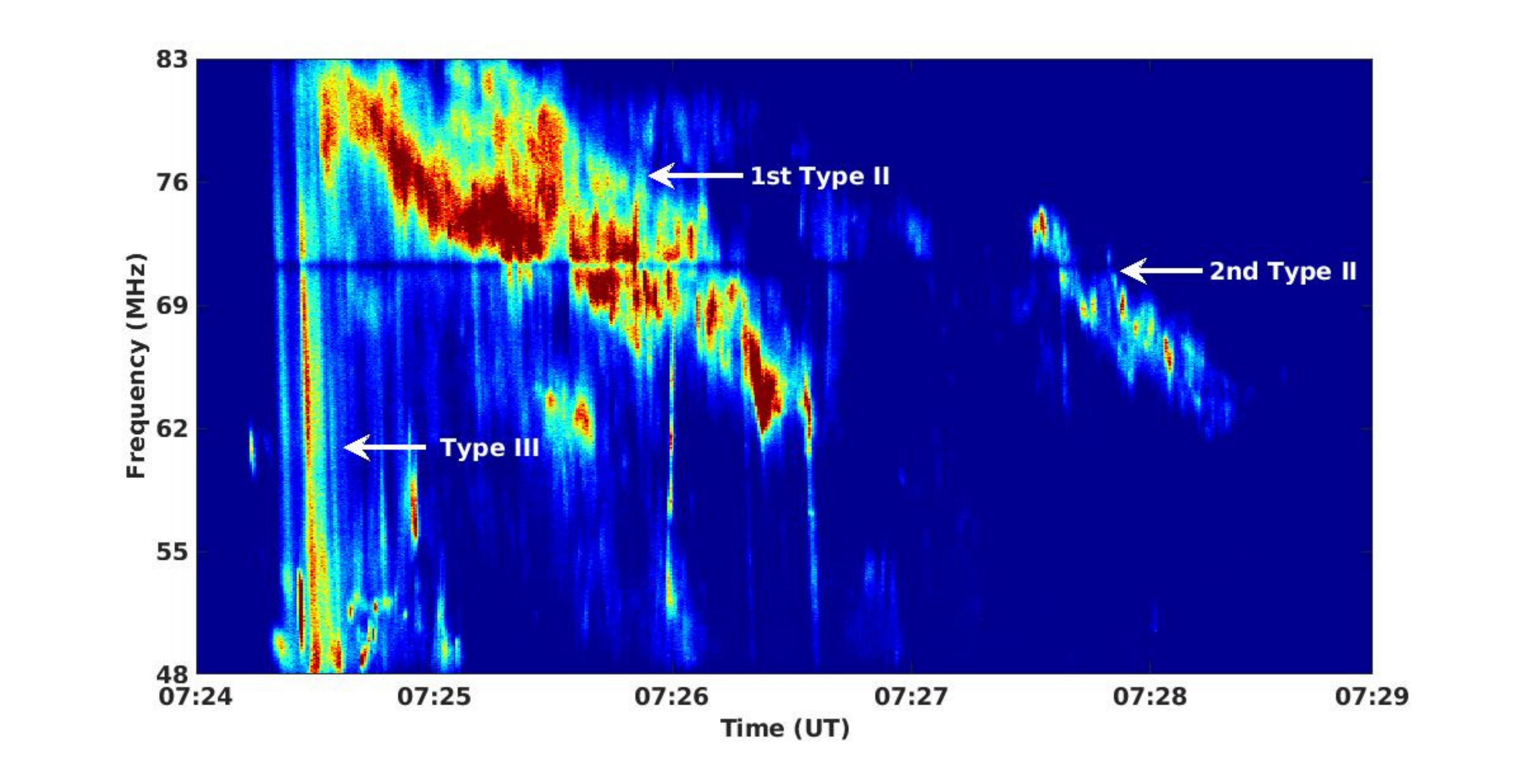}}
\caption{GAPS observations of transient radio emission from the solar corona on 2020 May 29. The fast drifting emission close to 
${\approx}$07:24:30\,UT is a type III burst. The relatively slow drifting emission during the intervals ${\approx}$07:24:30\,-\,07:26:30\,UT and
${\approx}$07:27:30\,-\,07:28:30\,UT are successive type II radio bursts.}
\label{fig:figure1}
\end{figure}

\begin{figure}[t!]
\centerline{\includegraphics[width=14cm]{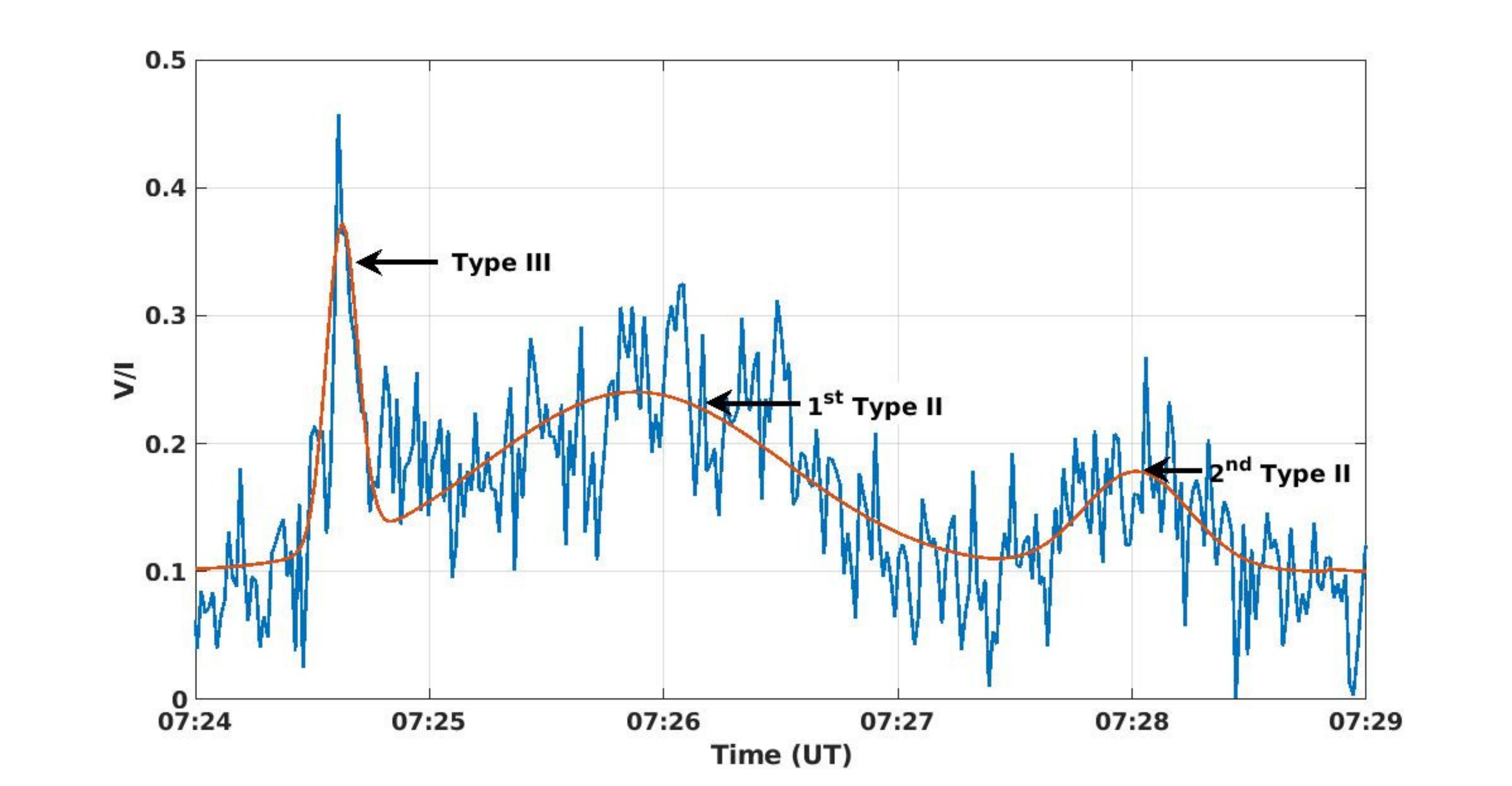}}
\caption{GRASP observations (65-70\,MHz) of the $dcp$ of the type III burst and successive type II bursts in Figure \ref{fig:figure1}. The red colour profile is the least squares fit to the data points.}
\label{fig:figure2}
\end{figure}

Figure \ref{fig:figure1} shows the GAPS observations of a type III burst followed by successive type II radio bursts from the solar corona on 2020 May 29. The overall bandwidths of the two type II bursts are limited. While the start frequency of the 1st type II burst seems to be 
${\gtrsim}$80\,MHz, its end frequency is ${\approx}$62\,MHz. Compared to this, the frequency range of the 2nd type II burst is 
${\approx}$75\,-\,62\,MHz. This is consistent with the statistical result that the start frequency of the 2nd type II burst in successive type II bursts is always lesser than that of the 1st type II burst \citep{Shanmugaraju2005,Subramanian2006}. The two type II bursts occurred during the time intervals ${\approx}$07:24:30\,-\,07:26:30\,UT and
${\approx}$07:27:30\,-\,07:28:30\,UT, respectively.  
They were associated with a M1.1 class GOES soft X-flare observed during the interval 
${\approx}$07:13\,-\,07:28\,UT. The maximum in the flare emission occurred at 
${\approx}$07:24\,UT. The flare location was at N32E89\footnote{https://www.lmsal.com/solarsoft/latest{\_}events{\_}archive/events{\_}summary/2020/05/29/gev{\_}20200529{\_}0718/index.html} near the east limb of the Sun. This indicates that the type II bursts in Figure \ref{fig:figure1} must be due to harmonic 
plasma emission since the corresponding fundamental (F) component from limb events as in the present case are likely to be occulted by the overlying corona and hence do not reach the observer. The directivity of F-component is also 
limited \citep[see e.g.][]{Nelson1985}. 
Figure \ref{fig:figure2} shows the $dcp$ obtained using the GRASP observations integrated over the frequency range ${\approx}$65\,-\,70\,MHz during the same time interval as in Figure \ref{fig:figure1}. The signal-to-noise ratio is poor 
due to the limited sensitivity of GRASP. Hence we used a least squares fit for the observed data points. It shows maxima in the $dcp$ near ${\approx}$07:24:30\,UT, ${\approx}$07:26\,UT, \& ${\approx}$07:28\,UT (indicated by arrow marks). These correspond to the type III, 1st and 2nd type II bursts in the GAPS dynamic spectrum in Figure \ref{fig:figure1}, respectively. The $dcp$ values of the aforementioned maxima (after subtracting the DC offset in the data) are 
${\approx}$0.27, 0.14, and 0.07, respectively. These are consistent with the earlier reports on $dcp$ for type III \& II bursts \citep{Dulk1980,Ramesh2010c,Sasikumar2013b,Hariharan2015,Anshu2017a,Anshu2019}.

\begin{figure}[t!]
\centerline{
\includegraphics[width=8cm]{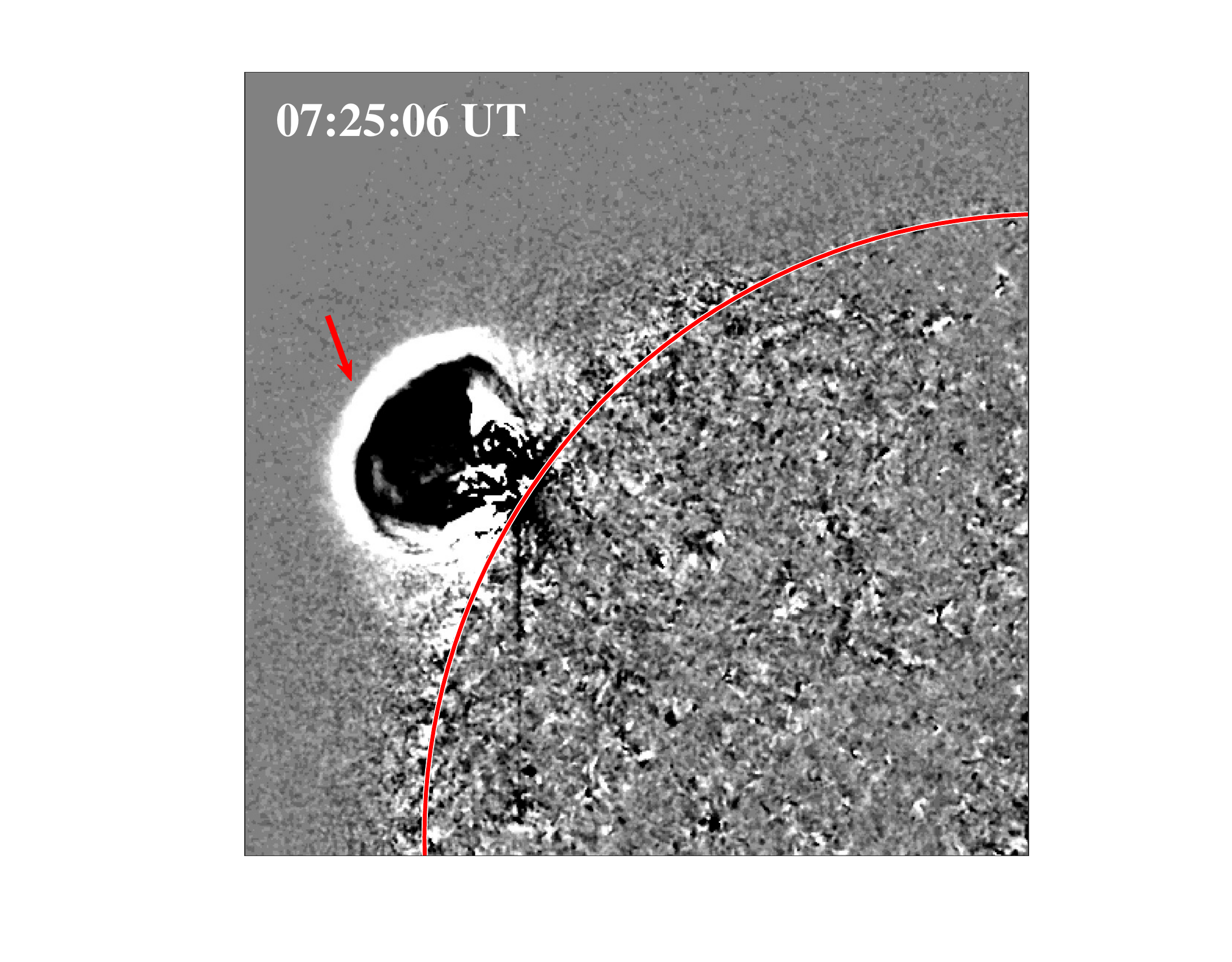}
\includegraphics[width=8cm]{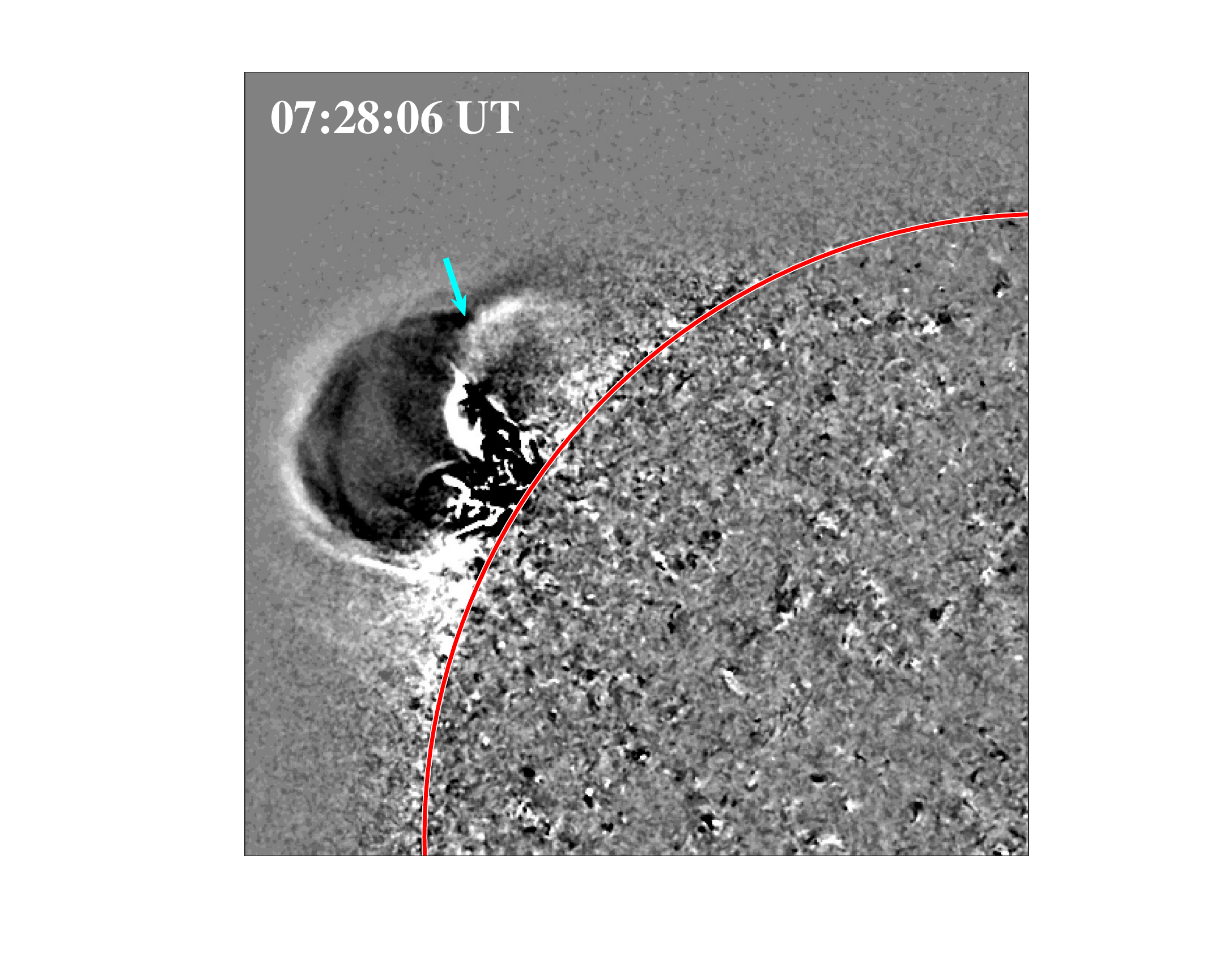}
}
\caption{LEFT: SDO/AIA-193{\AA} observations of the EUV flux rope (indicated by the red arrow) on 2020 May 29 at ${\approx}$07:25\,UT. Solar north is straight up and east is to the left. Image shown corresponds to the north-east quadrant of the Sun. The solar limb is indicated by the curved red line. Lower right corner in the image is the center of the Sun. RIGHT: Same as the image in the left panel, but observed at 
${\approx}$07:28\,UT. The cyan arrow indicates the rising EUV loop.}
\label{fig:figure3}
\end{figure}

An inspection of the SDO/AIA-193{\AA} running difference image obtained at 
${\approx}$07:25\,UT indicate that the 1st type II burst was associated with an EUV flux rope like structure (indicated by the red arrow in the left panel of Figure \ref{fig:figure3}) which propagated outwards from the same location as the flare mentioned above. Its position angle
(PA, measured counter-clockwise from the solar north) is ${\approx}50^{\arcdeg}$, and estimated linear speed is ${\approx}$477\,km/s in the SDO/AIA-193{\AA} field-of-view (FoV). The estimated speed of the MHD shock associated with the two bursts 
is ${\approx}$506${\pm}$33\,km/s according to the commonly used $N_{e}$ models for the solar corona \citep{Baumbach1937,Allen1947,Newkirk1961}. We used a density multiplier of 0.5 in the aforesaid models in order to match the speed of the EUV disturbance mentioned above. Since the present observations are close to the sunspot minimum period, use of the above density multiplier is 
justified \citep[see e.g.][]{Newkirk1967,Ramesh2020b}.
Note that the shock speeds obtained using other $N_{e}$ models were different despite attempts with different density multipliers. 
The leading edge (LE) of the flux rope was at $r{\approx}$1.29$R_{\odot}$ at ${\approx}$07:25\,UT when the 
1st type burst in Figure \ref{fig:figure1} was observed near 75\,MHz. According to the $N_{e}$ models mentioned above,
the plasma level corresponding 
to 37.5\,MHz plasma level (F-component) should be at 
$r{\approx}1.35{\pm}0.01R_{\odot}$. 
This is reasonable considering that low-frequency radio observations during the recent sunspot minimum period in 2019 indicate that the same plasma level in the background corona should be at $r{\approx}1.24R_{\odot}$ \citep[see e.g.][]{Ramesh2020b}.
The shock and the type
II burst are expected to be located ahead of the associated propagating disturbance, and the shock front, respectively. \cite{Gopalswamy2012} showed that for a propagating coronal disturbance like the EUV flux rope with LE at 
$r{\approx}$1.30$R_{\odot}$, the associated shock could be ahead by 
${\approx}$0.15$R_{\odot}$ (shock standoff distance). 
According to the statistical results of \cite{Suresh2016},
the standoff distance should be 
$0.16{\pm}0.1R_{\odot}$ near $r{\approx}1.3R_{\odot}$. Similar statistical work by \cite{Kim2012} indicates ${\approx}$0.2$R_{\odot}$ at the same distance. 
The expected locations of the type II burst (i.e. 37.5\,MHz plasma level) 
and the flux rope LE in the present case correspond to a shock standoff distance of 
${\approx}0.06R_{\odot}$. This is consistent with the aforementioned results. 
Hence we believe that
the 1st type II burst is due to the LE of the EUV flux rope in the left panel of Figure \ref{fig:figure3}. Due to technical reasons, we could not have coordinated imaging observations with the Gauribidanur radioheliograph \citep{Ramesh2014} for the present event to verify the location of the burst. Similar observations were not available elsewhere also. According to the SOHO/LASCO CME catalog, a CME was observed on 2020 May 29 at 
${\approx}$08:00\,UT with LE at $r{\approx}$3.1$R_{\odot}$. Its 
measurement position angle (MPA)
and angular width were 
${\approx}63^{\arcdeg}$ and ${\approx}37^{\arcdeg}$, 
respectively\footnote{https://cdaw.gsfc.nasa.gov/CME{\_}list/UNIVERSAL/2020{\_}05/yht/20200529.080005.w037n.v0337.p069g.yht}. The narrow bandwidth of the type II bursts is reasonably consistent with the latter \citep[see e.g.][]{Ramesh2022a}. The MPA of the LE was  
${\approx}75^{\arcdeg}$ at $r{\approx}5.8R_{\odot}$. 
The CME had a linear speed of ${\approx}$337\,km/s and deceleration of 
${\approx}$\,-13.2$\rm m/s^{2}$ in the SOHO-LASCO FoV. But, its initial speed in the range
$r{\approx}$1-2$R_{\odot}$ as per the quadratic fit to its height-time measurements is 
${\approx}$420\,km/s\footnote{https://cdaw.gsfc.nasa.gov/CME{\_}list/UNIVERSAL/2020{\_}05/htpng/20200529.080005.p069g.htp.html}. This is close to the propagation speed of the aforementioned EUV flux rope (i.e. ${\approx}$477\,km/s) in the present case. So, we think that the EUV flux rope in Figure \ref{fig:figure3} is the near-Sun signature of the CME.

The SDO/AIA-193{\AA} running difference image obtained at ${\approx}$07:28\,UT shows
an upward rising coronal loop at $r{\approx}$1.21$R_{\odot}$ and 
PA${\approx}40^{\arcdeg}$ near the northern flank of the same flux rope associated with the 1st type II burst (see cyan arrow in the right panel of Figure \ref{fig:figure3}).
The onset time of the 2nd type II burst in Figure \ref{fig:figure1} (${\approx}$07:27:30\,UT) at 
${\approx}$75\,MHz 
was close to the aforementioned epoch.
Since the plasma layer of the 
F-component (37.5\,MHz) of the burst 
is expected to be at $r{\approx}1.35{\pm}0.01R_{\odot}$, 
the shock standoff distance in this case is ${\approx}0.14R_{\odot}$.
This is within the range 
of the similar values mentioned in the literature (see previous paragraph). 
No other CMEs or propagating coronal disturbances were observed during the time interval between the two bursts in Figure \ref{fig:figure1}. Any possibility of association between the 2nd type II burst and X-ray flare mentioned earlier is also minimal since the latter had almost ended when the burst was observed \citep[see e.g.][]{Clasen2002,Ramesh2010b}. 
We further find from the different observations that:
(i) the temporal correlation between the onsets of the 2nd type II burst  
and movement of the coronal loop
located near the flank of the EUV flux rope is similar to the 
the association between the 1st type burst and LE of the same EUV flux rope;
(ii) the time interval between the appearance of the EUV flux rope
and beginning of the coronal loop movement
(${\approx}$3.0\,min) at nearly the same location (the difference between the respective $r$ values is ${\approx}0.08R_{\odot}$ only) is approximately equal to the delay between the start times of the 1st and 2nd type II bursts 
(${\approx}$2.5\,min) at the same frequency, i.e. 75\,MHz.
 Considering all the above details, we think that the 
coronal loop motion mentioned above is due to interaction between the earlier erupted EUV flux rope and the adjacent loops, and this resulted in the 2nd type II burst at the flank of the flux rope/CME
\citep[see e.g.][]{Reiner2003,Cho2007a,Cho2011,Feng2012,Chen2014,Hariharan2015}. The very close temporal correspondence (${\lesssim}$30\,sec) between the 
onset of the coronal loop movement
at the flank of the EUV flux rope, and the 2nd type II burst strengthens this reasoning. Note that the statistical results of \cite{Cho2008} indicate a time difference of 
${\lesssim}$2\,min in the case of CME flank-streamer interaction and the start time of the associated type II bursts. The gradual southward tilt (towards the equator) of the CME
from PA${\approx}50^{\arcdeg}$ at
$r{\approx}1.3R_{\odot}$ to PA${\approx}63^{\arcdeg}$ at $r{\approx}3.1R_{\odot}$ and then 
PA${\approx}75^{\arcdeg}$ 
at $r{\approx}5.8R_{\odot}$ with time (see previous paragraph), hints at deflected propagation of the CME. 
This is another potential evidence for the aforesaid interaction with the coronal loop
particularly since the latter was at the northern flank of the CME/EUV flux rope and the CME deflection was towards the south \citep[see e.g.][]{Wang2011}. Since it is a limb event, the above changes in PA are expected to be free of projection effects.

\section{Analysis and Results} \label{sec:anares}

\begin{figure}[t!]
\centerline{
\includegraphics[width=7cm]{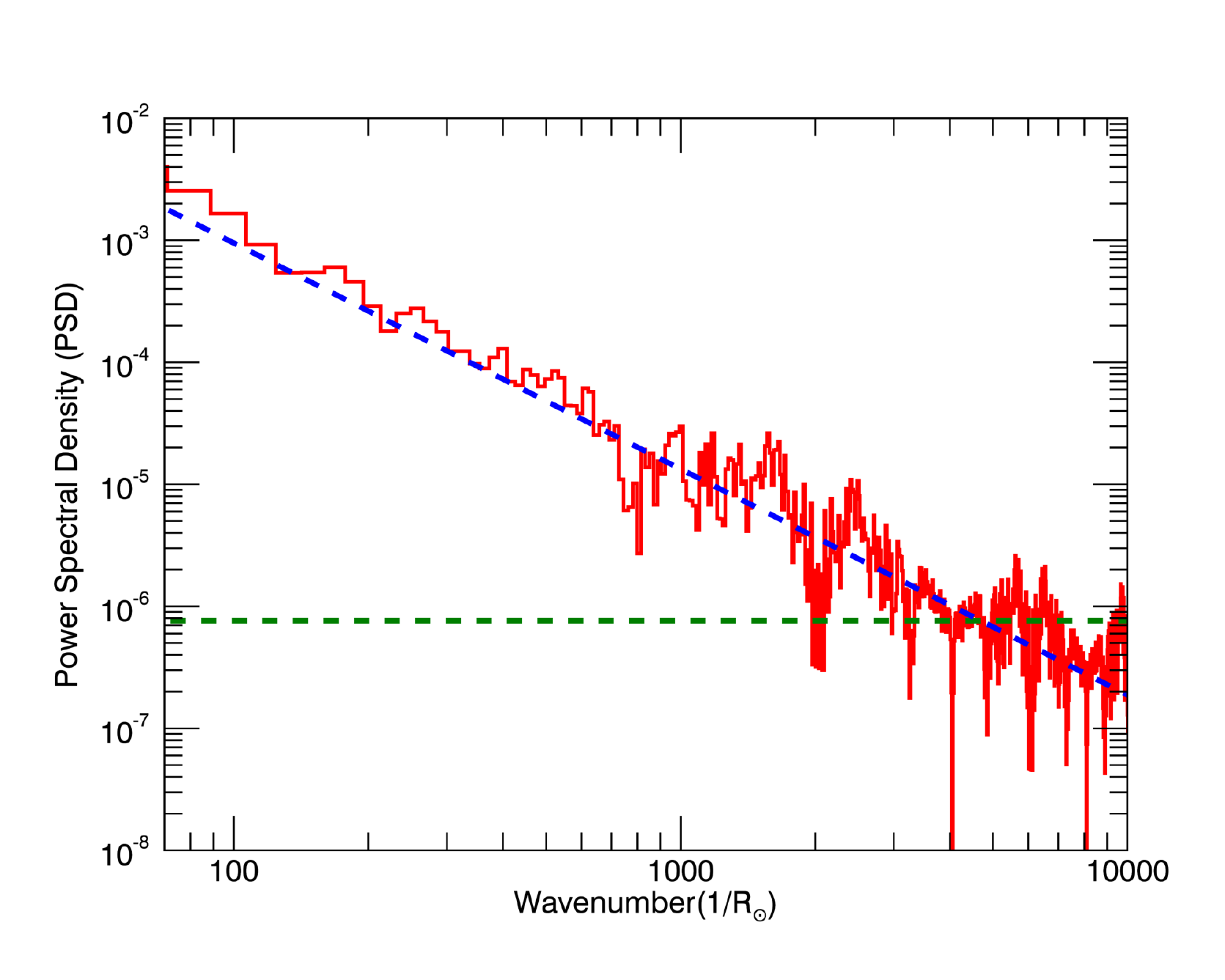}
\includegraphics[width=7cm]{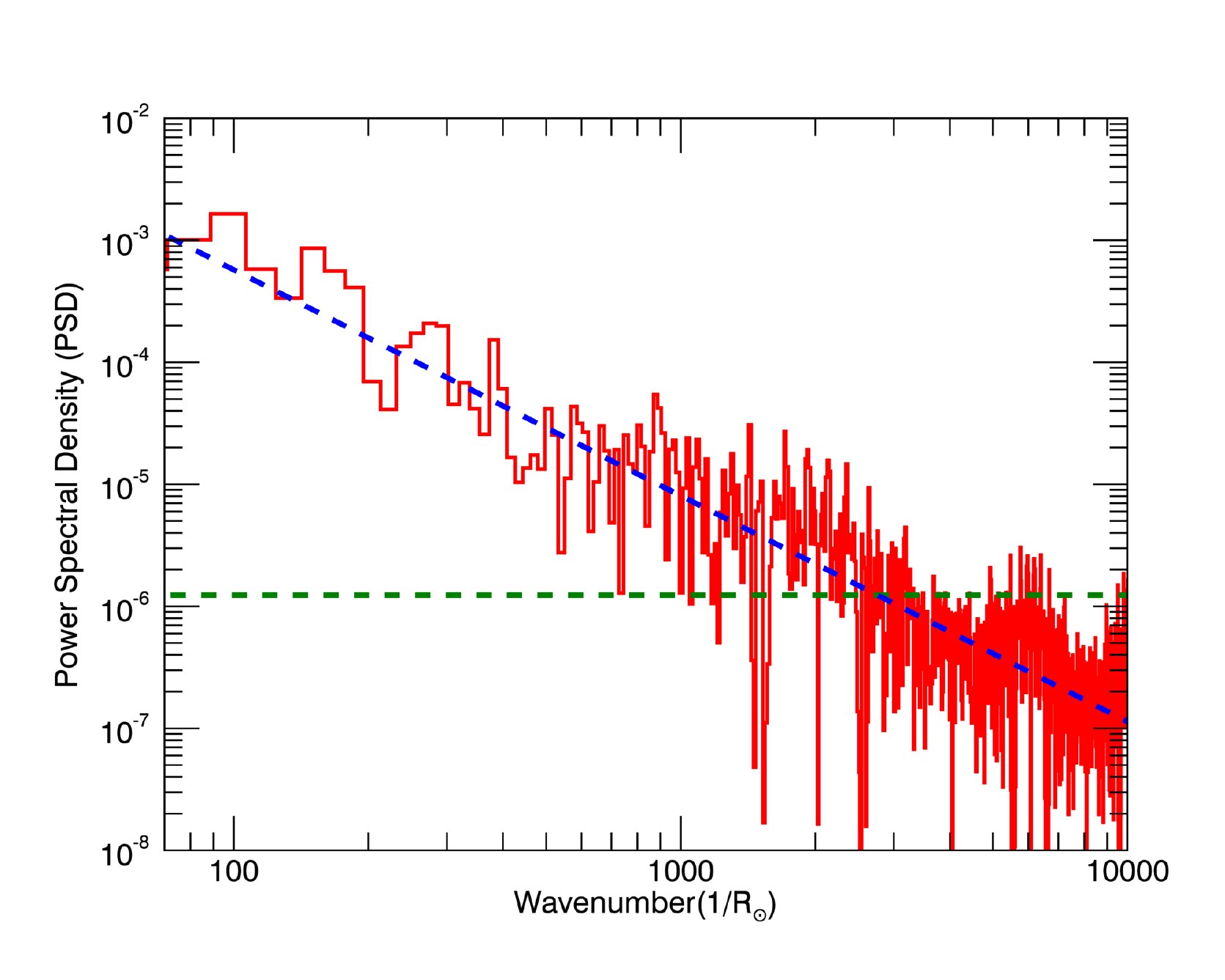}
}
\caption{LEFT: Power spectral density (PSD) corresponding to the 1st type II burst in Figure 
\ref{fig:figure1}. The inclined blue `dashed' line is the least squares fit to the estimated PSD. Its slope is ${\approx}$-1.85. The horizontal green `dashed' line indicates 5\% significance level. RIGHT: Same as the image in the left panel, but corresponds to the 2nd type II burst in Figure \ref{fig:figure1}. The unit for PSD in the present case is $\rm W^{2}m$.}
\label{fig:figure5}
\end{figure}

The fine structures and circular polarization exhibited by the the successive type II bursts in Figure \ref{fig:figure1} are related to the coronal density turbulence and magnetic field in the source region of the bursts, respectively. To infer the former, we estimated the PSD of the two bursts at different epochs as described in \cite{Chen2018} and \cite{Carley2021}.
The average PSD for the 1st and 2nd type II bursts are shown in Figure \ref{fig:figure5}. The slope is ${\approx}$-1.85 for both the bursts. 
This is same as the mean value at $r{\approx}2R_{\odot}$ reported by \cite{Carley2021} for typical non-successive type II bursts.
But the amplitude of the PSD corresponding to the 2nd type II burst 
($\rm {\approx}10^{-3}\,W^{2}m$) is 
${\approx}2{\times}$ lesser than that of the 1st type II burst 
($\rm {\approx}2{\times}10^{-3}\,W^{2}m$).
An inspection of Figure \ref{fig:figure2} indicates that the $dcp$ of the 2nd type II burst too is 
${\approx}2{\times}$ lesser compared to that of the 1st type II burst (${\approx}$0.07 \& 0.14, respectively). Note that for harmonic plasma emission and low values of $dcp$ as in the present case, $B{=}\frac{f_{p}{\times}dcp}{2.8\,a({\theta})}$ \citep{Melrose1980corrected,Willes1997}. $f_{p}$ (MHz) is the plasma frequency corresponding to fundamental component, and $a({\theta})$ depends on the angle of emission relative to the magnetic field direction. ${\theta}$ can be approximated to the heliographic longitude of the associated active region \citep{Dulk1980}. In the present case ${\theta}{\approx}89^{\arcdeg}$. For such near limb location, $a({\theta}){\gtrsim}$1 \citep{Dulk1978}. The GRASP observations of $dcp$ in Figure \ref{fig:figure2} correspond to harmonic plasma emission in the frequency range 
65-70\,MHz. This implies $f_{p}{\approx}$34\,MHz for both the 1st and 2nd type II bursts. Substituting
for $dcp$ in the above relation, we get $B{\lesssim}$1.8 \& 0.9\,G for the 1st and 2nd type II bursts, respectively. Note that the factor of two difference between the $B$ values of the two type II bursts in the present case is independent of $a({\theta})$. The latter is the same for both the 1st and 2nd type II bursts as they are associated with the LE and flank of the same flux rope, respectively, as discussed in Section 2 
(see Figure \ref{fig:figure3} also). 

According to the quasi-2D turbulence models, interaction between  
emerging and evolving loops in the `magnetic carpet' on the solar surface generates turbulence which is transferred into the corona and beyond \citep[see e.g.][] {Zank2021}. So, an increase or decrease in the magnetic field should lead to a corresponding change in turbulence \citep[see e.g.][]{Potherat2017}. Observations indicate that the level of turbulence in the solar wind varies with the Sun's magnetic field \citep{Janardhan2015,Sasi2019a}. For e.g., there was a 
${\approx}$17\% decrease in the global solar photospheric magnetic field 
during the period 1992-2018. The solar wind density turbulence 
decreased by ${\approx}$23\% in the same time interval. 
Therefore, the 
${\approx}$50\% decrease in the PSD of the 2nd type II burst (w.r.t. the 1st type II burst) in the present case must be due to its $B$ being lower by 
${\approx}$50\% compared to that of the 1st type II burst. The 1st and 2nd type II bursts are associated with the LE and flank of the EUV flux rope as mentioned in Section 2. Hence, the above results imply that the PSD and $B$ near the flux rope LE are ${\approx}2{\times}$ higher than the corresponding values near its flank. 
This could be because the LE is above the associated active region and the flank is outside the region. Note that the LE and flank of the flux rope are separated by ${\approx}10^{\circ}$ as mentioned earlier (see Section 2).
The results reported in \cite{Cho2007a} indicate 
similar ${\approx}2{\times}$ lower $B$ in the flank region of a CME as compared to its LE. Ray tracing calculations for polarization of thermal free-free radio emission from the solar corona with a density enhancement near the limb by \cite{Sastry2009} also indicate that the 
$dcp$ is lesser by ${\approx}2{\times}$ when $B$ in the enhancement is correspondingly reduced. 
Note that lower $B$ near the flank of the flux rope implies a lower Alf{\'v}en speed 
($v_{A}$) which favours shock formation in that region of the corona \citep[see e.g.][]{Jebaraj2021,Kouloumvakos2021}.

The power law in Figure \ref{fig:figure5} is in the wavenumber range ${\approx}$70\,-\,4500$R_{\odot}^{-1}$ (1st type II burst) and 
${\approx}$70\,-\,3000$R_{\odot}^{-1}$ (2nd type II burst), for PSD$>$5\% significance level. The corresponding ranges for the spatial scales in the turbulence (i.e. 2$\pi$/wavenumber) are 
${\approx}$62\,-\,1\,Mm and ${\approx}$62\,-\,1.5\,Mm, respectively.
A type II burst is generally expected to be located at the shock ahead of the associated propagating disturbance as mentioned before. So, the aforementioned turbulence is supposed to have existed in the coronal environment where the two type II bursts occurred in the present case. The upper limits are lesser than the outer scale of turbulence 
${\approx}278$\,Mm at $r{\approx}2R_{\odot}$ \citep{Bird2002,Mohan2021}.
The lower limits are greater than the dissipation scale of the turbulent density fluctuations at nearly the same $r$ \citep[see e.g.][]{Sasi2019b}. 
We would like to note here that the individual density irregularities reported earlier from the observed angular broadening of the Crab nebula at low radio frequencies due to its occultation by the solar corona are of size ${\sim}$1\,Mm \citep[see e.g.][]{Ramesh2001a}.

\section{Conclusions} \label{sum}

We have reported spectral and polarimeter observations of two weak, successive 
low-frequency (${\approx}$85\,-\,60\,MHz) type II radio bursts in the solar corona. 
Our results indicate that the 1st and 2nd type II bursts were generated by the leading edge of a flux rope / CME, and interaction of its flank with a neighbouring structure, respectively. The power spectral density and magnetic field strength of the 2nd type II burst (CME LE) are $2{\times}$ lesser than that of the 1st type II burst (CME flank) at the same $r$.
Considering that estimates of magnetic field strength from low-frequency radio observations of circularly polarized harmonic plasma emission as described in the present work are relatively easier to obtain, coordinated observations using 
ground- and space-based observing facilities with higher spectral and temporal resolutions \citep[see e.g.][]{Hariharan2016b} would be useful to understand the turbulence, magnetic field, etc. associated with the CMEs.
Such studies are expected to be important since there are reports that interplanetary CMEs with turbulent sheath region ahead of its LE drive stronger geomagnetic activity \citep{Kilpua2021}. Note that in the case of near-Sun observations, the diffuse structure observed ahead of the bright CME front near the Sun in some cases is regarded as the shock sheath \citep[see e.g.][]{Feng2013}. 
Moving further, we also found that the CME deflected away from radial direction, most likely after the aforesaid interaction.  
Such CMEs provide useful reference for space weather forecasting, especially for CME arrival and geoeffectiveness \citep{Wang2020}. This suggests a possible working hypothesis for a future research, i.e. whether sensitive observations of weak, successive coronal type II radio bursts as reported in the present work can be proxies for deflected CMEs close to the Sun. A larger data set of similar events is needed to verify this.
High cadence white-light observations in the range $1.05{\lesssim}r{\lesssim}3R_{\odot}$ (where the low-frequency coronal type II radio bursts as reported in the current work generally occur) with the Visible Emission Line Coronagraph \citep[VELC,][]{Singh2011} on board ADITYA-L1, the soon to be launched first Indian space solar mission, are expected to be helpful in this connection. 

We are grateful to Gauribidanur Observatory team for their help in the observations and upkeep of the facilities. Indrajit V. Barve, M. Rajesh, and K. P. Santosh are acknowledged for their contributions to the present work. The SOHO/LASCO CME catalog is generated and maintained at the CDAW Data Center by NASA and the Catholic University of America in cooperation with the Naval Research Laboratory. The SDO/AIA data are courtesy of the NASA/SDO and the AIA science teams. We thank the referee for his/her comments that helped to us to describe the results more clearly.




\end{document}